\documentclass[aps,pra,reprint,superscriptaddress,
longbibliography,floatfix]{revtex4-2}
\usepackage[T1]{fontenc}
\usepackage[utf8]{inputenc}
\usepackage{lmodern}
\usepackage{microtype}
\usepackage{amsmath,amssymb,amsthm,bm}
\usepackage{graphicx,booktabs}
\usepackage{tikz}
\usetikzlibrary{arrows.meta,positioning}
\definecolor{quantumblue}{RGB}{30,90,150}
\definecolor{classicalorange}{RGB}{175,85,15}
\usepackage{hyperref}
\hypersetup{colorlinks=true,linkcolor=blue,citecolor=blue,urlcolor=blue,
 pdftitle={Memory costs of quantum channel synthesis with a protected subsystem},pdfauthor={Kai Liu and Deguang Han}}
\newcommand{\M}{\mathrm M}
\newcommand{\C}{\mathbb C}
\newcommand{\id}{\operatorname{id}}
\newcommand{\Tr}{\operatorname{Tr}}
\newcommand{\Ad}{\operatorname{Ad}}
\newcommand{\spn}{\operatorname{span}}
\newcommand{\supp}{\operatorname{supp}}
\newcommand{\rank}{\operatorname{rank}}
\newcommand{\CA}{\mathfrak C_A}
\newcommand{\epsA}{\varepsilon^A}
\newcommand{\Dc}{\mathsf D}
\newtheoremstyle{apsstyle}
 {\topsep}{\topsep}{\normalfont}{\parindent}{\itshape}{.}{ }{}
\theoremstyle{apsstyle}
\newtheorem{theorem}{Theorem}
\newtheorem{proposition}{Proposition}

\newtheorem{corollary}{Corollary}

\begin{document}
\title{Intermediate memory costs of quantum channel synthesis with a protected subsystem}
\author{Kai Liu}
\author{Deguang Han}
\affiliation{Department of Mathematics, University of Central Florida,
Orlando, Florida, USA}

\begin{abstract}
A quantum memory can preserve a subsystem perfectly while discarding information needed by a later operation. We study the total intermediate dimension required for channel synthesis when the first stage protects a $b$-dimensional subsystem and has a prescribed noncommutative graph imposing full confusability on an $a$-dimensional control register. For a finite family of classically controlled branches, exact synthesis is possible if and only if the target graph contains the prescribed graph of the first stage. For every feasible family, we construct a single processor that implements all branches exactly. The processor uses pairwise nonorthogonal pure program states in a register of dimension at most $a$. A target graph may be realizable with a carrier of dimension $b$, while synthesizing a channel with that graph can require dimension $ab$. We also characterize the closure of the attainable channels, distinguishing exact synthesis from arbitrarily accurate approximation. Finally, for a family built from the trine communication task, a four-dimensional carrier permits exact synthesis with quantum programs under output depolarization. At the same total carrier dimension and without free shared randomness, classical programs have optimal normalized diamond error $(1-\eta)/4$, where $0\leq\eta\leq1$ is the depolarization strength. For $\eta<1$, the gap persists under sufficiently small perturbations of the target branches.
\end{abstract}

\maketitle

\section{Introduction}
Reducing the system passed between two or more stages of a quantum device forces information to be discarded. A subsystem may survive this reduction perfectly, yet a later operation can become impossible because information needed to choose its action has been lost. This raises a resource question: how much intermediate memory must be retained to implement a prescribed later process, and what error is unavoidable when the carrier is too small?

We study one use of a device with arbitrary quantum inputs and channel distinguishability assisted by a reference system. The resource is the total supported dimension crossing the intermediate cut, including classical flags; no additional memory, communication, shared randomness, or entanglement bypasses it. Related settings include memory costs of quantum networks \cite{ChiribellaNetworks,BisioMemory}, compression that preserves measurements \cite{BluhmCompression}, and dimension bottlenecks \cite{SekatskiBottleneck}. The bottleneck cost in Ref.~\cite{SekatskiBottleneck} allows free classical side information. Our cost includes those classical levels and imposes an exact confusability structure on the first channel.

For a channel $\Phi$ with Kraus operators $V_i$, its noncommutative graph is $S_\Phi=\spn\{V_i^*V_j\}$ \cite{DSW,Stahlke}. This operator system records which input states can remain perfectly distinguishable, without specifying the complete channel. We take an input $E\otimes F$ and impose
\begin{equation}
 S_\Phi=A:=\M_a\otimes\C I_b,
 \qquad \dim E=a,\quad \dim F=b.
 \label{eq:anchor}
\end{equation}
The weaker condition $S_\Phi\subseteq A$ already protects $F$ through the first stage \cite{NielsenPoulin,KribsSpekkens}, it also permits an identity encoder that retains the entire input. Equality additionally requires the Kraus products on $E$ to span $\M_a$. For controlled targets, this restriction has a concrete programming interpretation: the later stage must implement the selected branch using retained program states with pairwise nonorthogonal supports (Proposition~\ref{thm:programs}). This excludes a perfectly distinguishable record of the control label while allowing partial control information to survive. Throughout, protection and the graph constraint are exact, with no prescribed noise strength or uniform positive lower bound on program overlaps.

Established subsystem structure identifies every supported first stage, up to an output unitary, with $\Lambda\otimes\id_F$, where $\Lambda:E\to G$ is a channel \cite{NielsenPoulin,KribsSpekkens}. The intermediate carrier is $G\otimes F$: $F$ carries the protected data and $G$ retains information that the later operation may need. The second stage may act jointly on both registers and need not preserve $F$. The dimension of $G$ measures the memory required for dependence on $E$.

Our central synthesis result concerns a finite family of branch channels $\mathcal N_s$ selected by a classical label in $E$. Exact realization is possible if and only if the final graph contains $A$ (Theorem~\ref{prop:controlledfeasible}). Equivalently, the identity on $F$ belongs to every cross-Kraus span of two branches. When this condition holds, one processor and at most $a$ program dimensions suffice. The construction embeds all branch dilations into a common environment, so the pairwise conditions produce a simultaneous realization of the whole family. The retained states serve as programs with pairwise nonorthogonal supports in deterministic channel programming \cite{NielsenChuang,HilleryMaps,Heinosaari}. Cross-Kraus spans also occur in criteria for perfect discrimination of quantum operations \cite{DuanFengYing}. For two branches, the synthesis criterion gives the complete exact carrier cost: $b$, $2b$, or impossibility. A pair of amplitude damping branches illustrates an exact separation: a quantum program realizes the process, whereas every classical program satisfying the same overlap requirement fails.

A complementary result concerns realizing only the final graph. Extending the quantum complexity of a single graph defined in Ref.~\cite{LPT}, we minimize the intermediate dimension under both endpoint graph constraints. For $T\supseteq A$, the minimum equals $b$ precisely when
\begin{equation}
 T=\M_a\otimes R
 \label{eq:intro-split}
\end{equation} 
for an operator system $R$. Equivalently, $T$ is a bimodule over $\M_a\otimes\C I_b$, a structure also used in quantum relations and Morita equivalence \cite{Weaver,MoritaGraphs}. At carrier dimension $b$, the first stage must discard $E$, which forces this \emph{split} form. We give explicit nonsplit targets whose exact cost is $2b$. In contrast, specifying the entire channel can require dimension $ab$ even when its graph costs only $b$.

For approximate synthesis, we characterize the closure of the attainable processes and bound the error caused by discarding $E$. For controlled channels at dimension $b$, the optimal error is half the diamond norm radius of the branch family, with a semidefinite formulation. Two branches give one quarter of their diamond distance. A binary random unitary family exhibits zero infimum error without an exact realization. A Pauli family with three branches, built from an established trine communication task \cite{HeinosaariCommunication}, has costs $2$, $4$, and $6$ for graph realization, quantum synthesis, and classical synthesis, respectively. Matching error bounds quantify the quantum advantage at smaller carriers and its dependence on output noise. The advantage also persists under arbitrary sufficiently small perturbations of the branch channels.

Sections~\ref{sec:setup}--\ref{sec:process} develop the model and graph costs, and Sec.~\ref{sec:controlled} discusses controlled synthesis and its errors. Detailed proofs appear in the appendices.

\section{The sequential realization problem}
\label{sec:setup}

\subsection{Graphs and the intermediate carrier}
All Hilbert spaces are finite dimensional, and all channels are completely positive and trace preserving. We denote the identity channel by $\id$. For a channel $\Phi:\M_n\to\M_k$ with Kraus operators $V_i$, its noncommutative graph is
\begin{equation}
 S_\Phi=\spn\{V_i^*V_j:i,j\}\subseteq\M_n.
 \label{eq:graph}
\end{equation}
The graph is an operator space on the \emph{input}, even when the channel has a different output dimension. It is independent of the Kraus representation. It is an \emph{operator system}: a complex linear space containing the identity and closed under adjoints. Two pure inputs $|\psi\rangle,|\varphi\rangle$ have orthogonal channel outputs precisely when $\langle\psi|X|\varphi\rangle=0$ for every $X\in S_\Phi$ \cite{DSW}. Every finite-dimensional operator system can be realized as a channel graph \cite{DSW,LPT}. Two limiting examples help fix the convention: an identity channel has graph $\C I_n$, whereas discarding the whole input has graph $\M_n$. A larger graph imposes more conditions on inputs that can remain perfectly distinguishable.

We now consider a pair of channels describing a sequential device at a chosen intermediate cut, where the postprocessing channel $\mathcal D$ may represent the composition of several subsequent stages:
\begin{equation}
 \M_n\xrightarrow{\ \Phi\ }\M_k
 \xrightarrow{\ \mathcal D\ }\M_\ell.
 \label{eq:sequential}
\end{equation}
Only output levels that can actually be occupied are counted. The carrier dimension $w(\Phi)$ is therefore defined by
\begin{equation}
 w(\Phi)=\rank\Phi(I_n).
 \label{eq:width}
\end{equation}
Compressing the first output to this support changes neither channel graph nor the realizable composites, provided the postprocessor is restricted accordingly. Full support identifies occupied output levels, but it does not imply the ability to transmit arbitrary states on that space. Fresh ancillary systems initialized locally at either stage may be used freely. Every system or classical record retained from the first stage and available to the second is included in the counted carrier. An environment introduced locally to dilate the downstream processor therefore does not contribute to the intermediate carrier dimension.

If $R=S_{\mathcal D}$, the Kraus products of a composite give
\begin{equation}
 S_{\mathcal D\circ\Phi}
 =\spn\{V_i^*XV_j:X\in R,\ i,j\}.
 \label{eq:pullback}
\end{equation}
Since $I_k\in R$, postprocessing can only enlarge the graph: $S_\Phi\subseteq S_{\mathcal D\circ\Phi}$. This standard pullback relation underlies descriptions of noncommutative graphs in terms of channels \cite{Stahlke,Weaver}.

For operator systems $S\subseteq T\subseteq\M_n$, define the \emph{intermediate graph complexity} by
\begin{equation}
 \gamma_{\mathrm{int}}(S,T)=\min_{\Phi,\mathcal D}
 \{w(\Phi):S_\Phi=S,\ S_{\mathcal D\circ\Phi}=T\}.
 \label{eq:graphcost}
\end{equation}
The final output dimension is unrestricted. In contrast, the \emph{quantum complexity} of a single graph $S$ \cite{LPT} is
\begin{equation}
 \gamma(S)=\min_{S_\Phi=S}w(\Phi).
 \label{eq:gamma}
\end{equation}
Consequently, $\gamma_{\mathrm{int}}(S,T)\ge\gamma(S)$ and $\gamma_{\mathrm{int}}(S,S)=\gamma(S)$. The quantity with two endpoint constraints asks how much space is needed to realize $S$ while leaving enough information for the later graph $T$. The minimization is always feasible, and we later give a dimension bound after specializing to the protected subsystem.

\begin{figure}[t]
\centering
\begin{tikzpicture}[x=1cm,y=1cm,>=Latex,font=\small]
 \node at (-0.15,0.52) {$E$};
 \node at (-0.15,-0.52) {$F$};
 \draw[->] (0.08,0.52)--(0.6,0.52);
 \draw (0.6,0.17) rectangle (1.5,0.87);
 \node at (1.05,0.52) {$\Lambda$};
 \draw[->] (1.5,0.52)--(3.55,0.52);
 \node[above] at (2.15,0.52) {$G$};
 \draw[->] (0.08,-0.52)--(3.55,-0.52);
 \node[below] at (2.15,-0.52) {$F$};
 \draw (3.55,-0.92) rectangle (4.45,0.92);
 \node at (4,0) {$\mathcal D$};
 \draw[->] (4.45,0)--(5.25,0);
 \node[right] at (5.25,0) {$L$};
 \draw[dashed] (2.75,-1.12)--(2.75,1.12);
 \node[below,align=center] at (2.75,-1.13) {carrier dimension $bc$};
\end{tikzpicture}
\caption{Supported normal form of the sequential device. The graph $A$ of the first stage protects $F$ and requires $S_\Lambda=\M_a$. The cut carries $F$ and a retained register $G$, all information available to the second stage crosses this cut. Setting $c=1$ discards $E$ completely.}
\label{fig:carrier}
\end{figure}
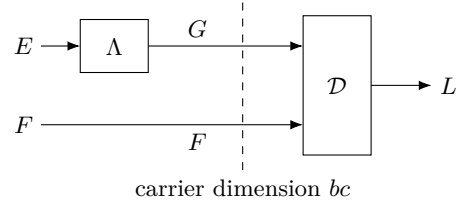

\subsection{What protection forces the carrier to retain}

The subsystem correction condition states that $F$ is correctable whenever $S_\Phi\subseteq A$ \cite{NielsenPoulin}. On the supported output, the corresponding structure for unitary recovery \cite{KribsSpekkens} takes the form
\begin{equation}
 U\Phi(X)U^*=(\Lambda\otimes\id_F)(X),
 \qquad K\cong G\otimes F.
 \label{eq:normal}
\end{equation}
Here $\Lambda:\M_a\to\M_c$ is a channel, $c=\dim G$, and $U$ is an output unitary. Moreover,
\begin{equation}
 S_\Phi=S_\Lambda\otimes\C I_b,
 \qquad w(\Phi)=bc.
 \label{eq:normalgraph}
\end{equation}
We give a direct proof that identifies the output support in Appendix~\ref{app:factor}. The output change of basis $U$ can be absorbed into the postprocessor, so the diagram represents every allowed device. Equality $S_\Phi=A$ is equivalent to $S_\Lambda=\M_a$, and full output support is ensured separately by compression.

There is also an environmental interpretation. Correctability of $F$ means that $F$ is private for the complementary channel describing the inaccessible environment \cite{KretschmannPrivacy}. The register $G$ in Fig.~\ref{fig:carrier} is different, it is part of the accessible output and may carry information needed by the second stage.

Every supported dimension is a multiple of $b$. At the smallest value, $c=1$, the only possible $\Lambda$ is the trace map, so
\begin{equation}
 w(\Phi)=b\quad\Longrightarrow\quad
 \Phi\simeq\Tr_E.
 \label{eq:minimumtrace}
\end{equation}
The second stage then has no access to $E$, which forces the split form in Eq.~\eqref{eq:intro-split}.

\subsection{A two-qubit example}
\label{sec:earlyexample}
Suppose $E$ is a control qubit and $F$ is a data qubit. The desired process discards control coherences and applies Pauli $Z$ to $F$ with probability $3/4$ for label $0$, but with probability $1/4$ for label $1$. The first stage can leave $F$ untouched and encode the control into the two distributions
\begin{equation}
 \rho_0=\operatorname{diag}(1/4,3/4),\qquad
 \rho_1=\operatorname{diag}(3/4,1/4).
 \label{eq:earlyprogram}
\end{equation}
The second stage reads the retained bit, applies $I$ or $Z$, and discards that bit. Because both distributions have full support, this encoder has the required full graph on $E$. The intermediate system contains two qubits, $G\otimes F$, with total dimension four. The additional qubit carries only classical information in this example.

Using only the protected qubit would discard $E$ by Eq.~\eqref{eq:minimumtrace}. The later operation would then be identical for both labels. Yet, for the same input $|+\rangle$ on $F$, the desired Pauli-$X$ expectations are $-1/2$ and $+1/2$. One protected qubit therefore cannot realize the target exactly. Section~\ref{sec:example} returns to this example and shows that its optimal one-qubit error, in the normalized diamond distance defined below, is $1/4$.

\section{Which downstream graphs need an extra register?}
\label{sec:graphs}
The two-qubit example fixes a complete target channel. We first consider the less restrictive task of reproducing only its graph, and returning to channel synthesis in Sec.~\ref{sec:process}. The minimum carrier dimension for $A$ is $b$, and fixing the downstream graph can force the device to retain more.

\begin{theorem}[Minimum dimension and the split criterion]
\label{thm:split}
Let $A=\M_a\otimes\C I_b$ and $A\subseteq T\subseteq\M_{ab}$ be an operator system. Then
\begin{equation}
 \gamma_{\mathrm{int}}(A,T)=b
 \quad\Longleftrightarrow\quad
 T=\M_a\otimes R
 \label{eq:split}
\end{equation}
for an operator system $R\subseteq\M_b$. Equivalently, $T$ is closed under left and right multiplication by $A$. In general,
\begin{equation}
 \gamma_{\mathrm{int}}(A,T)\in\{b,2b,\ldots,(a+1)b\}.
 \label{eq:levels}
\end{equation}
Further, for every $a,b\ge2$, there are indeed nonsplit targets with $\gamma_{\mathrm{int}}(A,T)=2b$.
\end{theorem}

The multiples of $b$ follow from the structure of the protected subsystem. The later constructions in Eqs.~\eqref{eq:sharpT}--\eqref{eq:sharpR} attains the nonsplit lower bound with a classical retained register with two levels.

At dimension $b$, Eq.~\eqref{eq:minimumtrace} gives
\begin{equation}
 S_{\mathcal D\circ\Tr_E}=\M_a\otimes S_{\mathcal D}.
 \label{eq:tracegraph}
\end{equation}
Every split target is obtained this way because every $R$ is a channel graph. All other allowed dimensions are multiples of $b$. To obtain the upper bound, one can keep either a minimum-dimensional realization of the first graph or the unchanged input in two orthogonal, classically flagged sectors. More generally, this construction gives
\begin{equation}
 \gamma_{\mathrm{int}}(S,T)\le\gamma(S)+n,\qquad S\subseteq T\subseteq\M_n.
 \label{eq:universal}
\end{equation}
The second sector can generate $T$ while the first preserves $S\subseteq T$; both sectors count toward the carrier size. With $S=A$, this yields $(a+1)b$ and proves feasibility. Appendix~\ref{app:graph} supplies the explicit channels. Closure under multiplication by $A$ is the standard matrix bimodule characterization of the tensor form \cite{Weaver,MoritaGraphs}. Appendix~\ref{app:graph} includes the proof using matrix units. The flagged construction uses the direct sum mechanism underlying quantum complexity \cite{LPT}.

To see that the lower bound is sharp, choose a nonscalar Hermitian $Z\in\M_b$ and define
\begin{equation}
 T_Z=A+\C\bigl(|1\rangle\langle1|_E\otimes Z\bigr).
 \label{eq:sharpT}
\end{equation}
Only one input sector has acquired the additional graph direction $Z$. Multiplication by matrix units on $E$ would spread that direction to all sectors, so $T_Z$ is not split. Equivalently, its dimension $a^2+1$ cannot equal $a^2\dim R$.

A retained register with two levels suffices. For $0<p<1$, let the channel on $E$ have Kraus operators
\begin{align}
 C_0&=\sqrt p\,|1\rangle_G\langle1|_E,\nonumber\\
 C_1&=\sqrt{1-p}\,|0\rangle_G\langle1|_E,\nonumber\\
 C_r&=|0\rangle_G\langle r|_E\quad(2\le r\le a).
 \label{eq:sharpC}
\end{align}
Their products span $\M_a$, and both output levels are occupied. Choose a postprocessor with graph
\begin{equation}
 R_Z=\spn\{I_{2b},|1\rangle\langle1|_G\otimes Z\}.
 \label{eq:sharpR}
\end{equation}
In Eq.~\eqref{eq:pullback}, the identity generates $A$, while only $C_0$ meets the second graph direction. Thus the composite graph is exactly $T_Z$ and $\gamma_{\mathrm{int}}(A,T_Z)=2b$. For two input qubits, adding one operator direction to the graph doubles the minimum physical carrier from two to four levels.

The sharp example also shows why a larger graph need not require a larger carrier. Although
\begin{equation*}
 T_Z\subsetneq\M_a\otimes\spn\{I_b,Z\},
\end{equation*}
the smaller, nonsplit target costs $2b$ and the larger, split target costs $b$. Requiring an exact graph can make selected directions costly to generate without also generating others. Nonmonotonicity already occurs for the quantum complexity $\gamma$ of a single graph \cite{LPT}, here it concerns the downstream target with the graph of the first stage held fixed.

\section{From graph realization to process simulation}
\label{sec:process}
The graph realization problem specifies only the graph of the final channel, not the channel itself. To describe a specified process $\mathcal N:\M_{ab}\to\M_\ell$ through the sequential device in Eq.~\eqref{eq:sequential}, define the attainable class
\begin{equation}
 \CA(k)=\{\mathcal D\circ\Phi:S_\Phi=A,\ w(\Phi)\le k\},
 \label{eq:class}
\end{equation}
and its exact cost and approximation error by
\begin{align}
 m_A(\mathcal N)&=\min\{k:\mathcal N\in\CA(k)\},
 \label{eq:processcost}\\
 \epsA_k(\mathcal N)&=\frac12\inf_{\mathcal M\in\CA(k)}
       \|\mathcal N-\mathcal M\|_\diamond.
 \label{eq:error}
\end{align}
We adopt the convention that a minimum or infimum over an empty set equals $+\infty$. The diamond norm gives the operational distance for one use of a channel, allowing a reference system whose entanglement with the input can improve discrimination \cite{Sacchi,WatrousSDP,PianiWatrous}. The factor $1/2$ normalizes the error to lie between zero and one whenever an admissible device exists. The cost and error are not determined by the final graph alone. For $k\ge b$, Eq.~\eqref{eq:normalgraph} implies
\begin{equation}
 \CA(k)=\CA\bigl(b\lfloor k/b\rfloor\bigr).
 \label{eq:stair}
\end{equation}
For $k<b$, the class is empty and no device is admissible.

\subsection{A general test for discarding the input register}
\label{sec:erase}
At dimension $b$, the first stage discards $E$ completely. This yields a criterion for exact synthesis at the minimum carrier dimension that applies to arbitrary target channels. Define the channel\begin{equation}
 \mathcal E_E(X)=\frac{I_E}{a}\otimes\Tr_E X.
 \label{eq:erase}
\end{equation}
It replaces $E$ by a fixed state while keeping the marginal on $F$. A target can use the minimum carrier exactly when this erasure leaves it unchanged, that is,
\begin{equation}
 m_A(\mathcal N)=b
 \quad\Longleftrightarrow\quad
 \mathcal N=\mathcal N\circ\mathcal E_E.
 \label{eq:erasetest}
\end{equation}
The corresponding operational error satisfies
\begin{equation}
 \frac{d_E(\mathcal N)}4\le\epsA_b(\mathcal N)
       \le\frac{d_E(\mathcal N)}2,
 \label{eq:erasebound}
\end{equation}
where $d_E(\mathcal N)=\|\mathcal N-\mathcal N\circ\mathcal E_E\|_\diamond$. This estimate from a contractive projection applies also to targets sensitive to coherences in $E$. Appendix~\ref{app:erase} gives the proof. In channel comparison, $2\epsA_b(\mathcal N)$ is the postprocessing deficiency of $\Tr_E$ relative to $\mathcal N$ \cite{Buscemi,JencovaGeneral}. We also compute the optimal approximation error explicitly for the binary random unitary and trine families in Secs.~\ref{sec:example} and~\ref{sec:trine}, respectively.

\subsection{An unbounded separation between graph and process costs}
A small depolarizing admixture makes the graph on $E$ full while preserving linear invertibility. Such a process cannot pass through a smaller matrix space, even though its graph has minimum cost.

\begin{proposition}[Separation between graph and process costs]
\label{prop:invertible}
Suppose $\mathcal N:\M_{ab}\to\M_{ab}$ is linearly invertible and $S_{\mathcal N}=A$. Then
\begin{equation}
 \gamma_{\mathrm{int}}(A,S_{\mathcal N})=b,\qquad m_A(\mathcal N)=ab.
 \label{eq:gap}
\end{equation}
For every $b\le k<ab$,
\begin{equation}
 \epsA_k(\mathcal N)\ge
 \frac{1}{2\|\mathcal N^{-1}\|_\diamond}.
 \label{eq:inverse}
\end{equation}
Here, the inverse is a linear map and need not be a channel.
\end{proposition}

Factoring through $\M_k$ bounds the linear rank by $k^2$. Since $\mathcal N$ is linearly invertible, its rank is $(ab)^2$, forcing $k\ge ab$, while choosing $\Phi=\mathcal N$ and $\mathcal D=\id$ attains equality. The error bound follows from a Hermitian kernel element of the map using a smaller carrier (Appendix~\ref{app:bottleneck}).

For a concrete family, take
\begin{align}
 \Omega_a(X)&=\Tr(X)I_a/a,\nonumber\\
 \mathcal P_{q,a}&=(1-q)\id+q\Omega_a,\qquad0<q<1,\nonumber\\
 \mathcal N_q&=\mathcal P_{q,a}\otimes\id_b.
 \label{eq:depol}
\end{align}
The nonzero depolarizing component gives $S_{\mathcal P_{q,a}}=\M_a$, while $q<1$ ensures invertibility. Thus the graph cost is $b$ but the process cost is $ab$, with ratio $a$.

If the dimension of the retained register is $c<a$, the map $\Lambda:\M_a\to\M_c$ has a nonzero Hermitian kernel element $H$. Trace preservation makes $H$ traceless. The first stage erases $H\otimes\tau$ for any state $\tau$ on $F$, whereas the target sends it to $(1-q)H\otimes\tau$. Therefore
\begin{equation}
 \epsA_k(\mathcal N_q)\ge\frac{1-q}{2},
 \qquad b\le k<ab.
 \label{eq:depolbound}
\end{equation}
Thus every smaller carrier incurs positive error for fixed $q<1$. The bound vanishes as $q\to1$, when $E$ is completely depolarized.

\subsection{Exact graph constraints and their closure}
\label{sec:closure}
At fixed input, carrier, and output dimensions, unrestricted factorizations into an encoder and a decoder form a compact set. Here the exact graph constraint $S_\Phi=A$ need not be preserved under limits since arbitrarily weak depolarizing noise on $E$ can enforce this constraint, even when the limiting channel does not satisfy it. The following proposition identifies exactly which limits this constraint excludes.

\begin{proposition}[Closure of the attainable processes]
\label{thm:closure}
For a fixed output $\M_\ell$ and an integer $c\ge1$, let
\begin{equation}
 \mathcal F_c=\left\{\mathcal D\circ(\Lambda\otimes\id_b):
 \begin{array}{l}
 \Lambda:\M_a\to\M_c,\\
 \mathcal D:\M_{cb}\to\M_\ell
 \end{array}
 \text{ channels}\right\}.
 \label{eq:relaxed}
\end{equation}
Unused output levels are allowed in this definition. Then
\begin{equation}
 \overline{\CA(cb)}=\mathcal F_c,
 \label{eq:closure}
\end{equation}
where closure is taken in diamond norm. Consequently,
\begin{equation}
 \epsA_{cb}(\mathcal N)=\frac12
 \min_{\Lambda,\mathcal D}
 \|\mathcal N-\mathcal D\circ(\Lambda\otimes\id_b)\|_\diamond.
 \label{eq:relaxedmin}
\end{equation}
\end{proposition}

Perturb any candidate first map by
\begin{equation}
 \Lambda_\delta=(1-\delta)\Lambda+
       \delta\Tr(\cdot)I_c/c,\qquad0<\delta<1.
 \label{eq:fulldense}
\end{equation}
The depolarizing part makes $S_{\Lambda_\delta}=\M_a$ and fills the output support. The normalized error introduced in the composite is at most $\delta$. Conversely, every constrained realization can be embedded in a $c$-dimensional retained register. Compactness of the finite-dimensional channel sets then proves the proposition. See Appendix~\ref{app:closure}.

The relaxed class $\mathcal F_c$ is exactly the class obtained by imposing only $S_\Phi\subseteq A$ at the same budget. Full confusability can therefore change exact attainability without increasing the infimum error, with protection alone, the optimum is always attained. Taking $c=a$, $\Lambda=\id_E$, and $\mathcal D=\mathcal N$ gives $\epsA_{ab}(\mathcal N)=0$ for every target, so nontrivial approximation occurs below the original input dimension. The infimum under the graph constraint need not be attained:
\begin{equation}
 \epsA_{cb}(\mathcal N)=0
 \quad\not\Longrightarrow\quad m_A(\mathcal N)\le cb.
 \label{eq:nonattain}
\end{equation}
Section~\ref{sec:example} gives an explicit example. For $c>1$ the optimization in Eq.~\eqref{eq:relaxedmin} is generally nonconvex in the encoder and decoder jointly, and the closure characterization does not provide an efficient global algorithm.

\section{Classically controlled processes and quantum programs}
\label{sec:controlled}

\subsection{Exact realization and nonorthogonal programs}
Fix a basis $|s\rangle$ of $E$ and let $X_{st}=(\langle s|\otimes I)X(|t\rangle\otimes I)$. A process that reads the classical label in $E$, discards its coherences, and applies a branch channel to $F$ has the form
\begin{equation}
 \mathcal N_\oplus(X)=\sum_{s=1}^a\mathcal N_s(X_{ss}),
 \qquad \mathcal N_s:\M_b\to\M_\ell.
 \label{eq:controlled}
\end{equation}
On label $s$, the retained register contains a state $\rho_s$ that programs $\mathcal N_s$ through one fixed processor. This is the established setting of programmable channels \cite{NielsenChuang,HilleryMaps,HilleryPrograms,Heinosaari}. The additional requirement here is the exact graph of the first stage. Write $\Dc(H)$ for the density operators on $H$.

\begin{proposition}[Programs for a controlled process]
\label{thm:programs}
The process in Eq.~\eqref{eq:controlled} belongs to $\CA(cb)$ if and only if there exist states $\rho_s\in\Dc(\C^c)$ and a channel $\mathcal P:\M_c\otimes\M_b\to\M_\ell$ such that
\begin{align}
 \mathcal P(\rho_s\otimes X)&=\mathcal N_s(X)
                 &&\text{for all }s,X,\label{eq:programs}\\
 \Tr(\rho_s\rho_t)&>0&&\text{for all }s,t.
 \label{eq:overlap}
\end{align}
Unused program dimensions may be removed.
\end{proposition}

The overlap condition excludes orthogonal program supports. For an admissible encoder, set $\rho_s=\Lambda(|s\rangle\langle s|)$. Its Kraus operators obey 
\begin{equation}
 \Tr(\rho_s\rho_t)=
 \sum_{i,j}|\langle s|C_i^*C_j|t\rangle|^2>0,
 \label{eq:overlapidentity}
\end{equation}
since $\spn\{C_i^*C_j\}=\M_a$. Conversely, the measure-and-prepare encoder $\Lambda(X)=\sum_s\langle s|X|s\rangle\rho_s$ has graph $\M_a$ exactly when Eq.~\eqref{eq:overlap} holds. Appendix~\ref{app:programs} gives the proof.

A \emph{classical retained register} has commuting program states. The encoder and processor are fixed, with local randomness but no shared random seed or entanglement outside the counted carrier. Dephasing the input label cannot increase error, so this restriction includes arbitrary classical output encoders for the controlled targets considered here. Write $m_A^{\rm cl}$ and $\varepsilon_k^{A,\rm cl}$ for the corresponding restrictions of Eqs.~\eqref{eq:processcost} and \eqref{eq:error}.

Exact feasibility can be decided without searching over programs or processors. Let $K_{s\mu}$ be any Kraus family for $\mathcal N_s$.

\begin{theorem}[Feasibility of controlled processes]
\label{prop:controlledfeasible}
For the channel in Eq.~\eqref{eq:controlled}, the following conditions are equivalent:
\begin{align}
 &m_A(\mathcal N_\oplus)<\infty,\nonumber\\
 &A\subseteq S_{\mathcal N_\oplus},\nonumber\\
 &I_b\in\spn\{K_{s\mu}^*K_{t\nu}:\mu,\nu\}
                 \quad\text{for every }s,t.
 \label{eq:controlledfeasible}
\end{align}
When these conditions hold, one fixed processor can implement all branches using linearly independent pure program states in $\C^a$ with pairwise nonzero overlaps. In particular, $m_A(\mathcal N_\oplus)\le ab$.
\end{theorem}

The test requires only solving finite linear systems to determine whether $I_b$ belongs to each cross-Kraus span. It suffices to check $s<t$, because trace preservation handles $s=t$ and adjoints handle reversed pairs. Necessity follows from graph monotonicity. Sufficiency also requires the pairwise solutions to fit one processor. Scaling them by a common small positive factor makes their block Gram matrix positive. Factoring this matrix, as in Appendix~\ref{app:feasible}, embeds all branch Stinespring isometries $V_s$ in a common environment so that $V_s^*V_t=\alpha I_b$ for every $s\ne t$, with one sufficiently small $\alpha>0$. The positive Gram matrix $(1-\alpha)I_a+\alpha\boldsymbol1\boldsymbol1^*$ then supplies the pure programs and the common processor.

The programming relation $V_s^*V_t=\langle p_s|p_t\rangle I_b$ \cite{HilleryMaps,ZimanBuzek} involves both program vectors and environmental dilations. Theorem~\ref{prop:controlledfeasible} expresses compatibility directly in terms of the cross-Kraus spans of the specified branches, and the joint dilation construction above makes the pairwise conditions sufficient for the whole family. It gives the carrier bound $ab$, but need not minimize the program dimension. Identity exclusion from a cross-Kraus span is also one of the conditions for perfect channel discrimination with finitely many queries \cite{DuanFengYing}, where inclusion in every such span characterizes synthesis under the graph constraint.

Programming bounds can also be obtained from fidelity constraints on program states and their induced channels, as shown by Tukiainen \emph{et al.} \cite{TukiainenFidelity}. For covariant channel families, Gschwendtner, Bluhm, and Winter~\cite{GschwendtnerCovariant} establish exact programming in finite dimension and determine optimal program dimensions when the group acts irreducibly on the channel input. The criterion above applies to a prescribed finite family without a covariance assumption, with pairwise nonorthogonal program supports required by the first stage graph.

If every branch has a positive definite Choi matrix, its Kraus operators span all maps from $F$ to $L$. Each cross-Kraus span is then $\M_b$, so the theorem applies. For any family, the perturbed branches $(1-\delta)\mathcal N_s+\delta\Tr(\cdot)\omega$, with $0<\delta<1$ and a positive definite density operator $\omega$ on $L$, have positive definite Choi matrices. Thus an arbitrarily weak common replacement admixture makes the perturbed family exactly feasible with at most $ab$ carrier dimensions, even if the original family is infeasible.

For two branches the theorem gives the complete exact cost. Put $\mathcal V_{01}=\spn\{K_{0\mu}^*K_{1\nu}\}$. Then
\begin{equation}
 m_A(\mathcal N_\oplus)=
 \begin{cases}
 b,&\mathcal N_0=\mathcal N_1,\\
 2b,&\mathcal N_0\ne\mathcal N_1,\ I_b\in\mathcal V_{01},\\
 +\infty,&I_b\notin\mathcal V_{01}.
 \end{cases}
 \label{eq:binaryexact}
\end{equation}
Dimension $b$ discards the label, so it requires identical branches, otherwise the theorem attains the next allowed dimension $2b$ whenever feasible. For distinct unitary channels, $\mathcal V_{01}=\C U_0^*U_1$ excludes $I_b$. This reproduces the obstruction from the no-programming theorem \cite{NielsenChuang,Heinosaari}, which shows that exact programs would need orthogonal supports, contrary to Eq.~\eqref{eq:overlap}.

\subsection{Approximation and the loss of control information}
For approximation, Proposition~\ref{thm:closure} removes the overlap requirement. The error for any program dimension is
\begin{equation}
 \epsA_{cb}(\mathcal N_\oplus)=\frac12
 \min_{\rho_1,\ldots,\rho_a,\mathcal P}
 \max_s\|\mathcal N_s-\mathcal P(\rho_s\otimes\cdot)\|_\diamond.
 \label{eq:programapprox}
\end{equation}

Dephasing $E$ before an approximate realization leaves the target unchanged and cannot increase the error. The resulting encoder prepares a program state for each label. The identity for the block norm in Appendix~\ref{app:programs} then identifies the channel error with the largest branch error, including for inputs assisted by a reference system.

For a fixed processor, optimizing the program states is a convex problem, as studied in Ref.~\cite{Banchi}. Equation~\eqref{eq:programapprox} also optimizes the processor, thus this joint problem is generally nonconvex.

At the minimum carrier size, the retained register has dimension one. All programs are identical, and a single channel $\mathcal D$ must approximate every branch. Equation~\eqref{eq:programapprox} becomes
\begin{equation}
 \epsA_b(\mathcal N_\oplus)
 =\frac12\min_{\mathcal D}\max_s
 \|\mathcal N_s-\mathcal D\|_\diamond,
 \label{eq:radius}
\end{equation}
where the minimum is taken over all channels $\mathcal D:\M_b\to\M_\ell$.
Thus the optimal decoder is a center of minimum radius of the branch channels in diamond norm.

If $d_{\max}=\max_{s,t}\|\mathcal N_s-\mathcal N_t\|_\diamond$, the triangle inequality and the average channel give
\begin{equation}
 \frac{d_{\max}}4\le\epsA_b(\mathcal N_\oplus)
 \le\frac{a-1}{2a}d_{\max}.
 \label{eq:radiusbounds}
\end{equation}
For two branches, the midpoint is optimal and
\begin{equation}
 \epsA_b(\mathcal N_\oplus)
 =\frac14\|\mathcal N_0-\mathcal N_1\|_\diamond.
 \label{eq:binary}
\end{equation}
For more than two branches, Eq.~\eqref{eq:radius} is a semidefinite optimization using the usual diamond norm constraints \cite{WatrousSDP}. Appendix~\ref{app:radius} records an explicit formulation.

Equation~\eqref{eq:binary} also has an operational interpretation. For two equally likely branches and a common optimized probe on $F$ and a reference system, the optimal discrimination probability is \cite{Sacchi}
\begin{equation}
 p_{\rm succ}^{\star}
 =\frac12+\frac14\|\mathcal N_0-\mathcal N_1\|_\diamond.
 \label{eq:success}
\end{equation}
The advantage over random guessing, $p_{\rm succ}^{\star}-1/2$, is therefore exactly $\epsA_b(\mathcal N_\oplus)$. A dimension-$b$ realization produces identical outputs for the two labels on that probe, so it loses this discrimination advantage completely. This comparison uses the same input on the data and reference systems for both labels, while the channel model itself still allows arbitrary correlated inputs on $E\otimes F$.

\subsection{A binary random unitary family with an exact relation between dimension and error}
\label{sec:example}
Let $a=2$, let $U$ be a unitary on $F$, write $\Ad_U(X)=UXU^*$, and consider
\begin{equation}
 \mathcal N_{p,U}(X)=\mathcal B_+(X_{00})+\mathcal B_-(X_{11}),
 \label{eq:example}\end{equation}
where $0\le p\le1$ and
\begin{equation}
 \mathcal B_\pm=\frac{1\mp p}{2}\id+\frac{1\pm p}{2}\Ad_U.
 \label{eq:branches}
\end{equation}
The two-qubit construction in Sec.~\ref{sec:earlyexample} is the case $U=Z$ and $p=1/2$. More generally, the parameter $p$ measures the contrast between the two control labels. At $p=0$ the branches coincide; at $p=1$ the labels select $\Ad_U$ and $\id$ deterministically. For $0<p<1$, a retained register with two levels realizes the channel exactly. Use the programs
\begin{align}
\rho_0&=\frac{1-p}{2}|0\rangle\langle0|
+\frac{1+p}{2}|1\rangle\langle1|,\nonumber\\
 \rho_1&=\frac{1+p}{2}|0\rangle\langle0|+\frac{1-p}{2}|1\rangle\langle1|,
 \label{eq:exampleprogram}
\end{align}
and let the processor apply $I$ or $U$ conditional on the program basis state, then discard the program. Their overlap is $(1-p^2)/2>0$, so Proposition~\ref{thm:programs} ensures the required graph. No program coherence is needed.

Equation~\eqref{eq:binary} gives
\begin{align}
 e(p,U):=\epsA_b(\mathcal N_{p,U})
 &=\frac p4\|\Ad_U-\id\|_\diamond\nonumber\\
 &=\frac p2\sqrt{1-\nu(U)^2},
 \label{eq:uniterror}\\
 \nu(U)&=\min_{\rho\in\Dc(F)}|\Tr(\rho U)|.
 \label{eq:nu}
\end{align}
The last equality is the established formula for discriminating unitary channels \cite{Acin}. An optimal decoder is $(\id+\Ad_U)/2$, independent of $p$. Combining the lower bound with the construction with two program levels and its limiting versions gives, for every $0\le p\le1$,
\begin{equation}
 \epsA_k(\mathcal N_{p,U})=
 \begin{cases}
 +\infty,& k<b,\\
 e(p,U),& b\le k<2b,\\
 0,& k\ge2b.
 \end{cases}
 \label{eq:profile}
\end{equation}
For nonscalar $U$, the corresponding \emph{exact} costs are
\begin{equation}
 m_A(\mathcal N_{p,U})=
 \begin{cases}
 b,&p=0,\\
 2b,&0<p<1,\\
 +\infty,&p=1.
 \end{cases}
 \label{eq:exactprofile}
\end{equation}
For nonscalar $U$ and $p=1$, the two branches are distinct unitary channels. Exact synthesis is prohibited by no-programming, but the admissible channels $\mathcal N_{p,U}$ with $p<1$ converge to it. The zero error in the last line of Eq.~\eqref{eq:profile} is then an unattained infimum. Table~\ref{tab:pauli} summarizes this distinction for a protected qubit. For scalar $U$, the branches coincide and the exact cost is always $b$.

\begin{table}[!hb]
\caption{Carrier requirements for the classically controlled Pauli-$Z$ family ($b=2$). All dimensions include the protected qubit. At $p=1$, the two-qubit error has infimum zero, but no admissible device attains it.}
\label{tab:pauli}
\centering
\begin{tabular}{@{}lccc@{}}
\toprule
Control & Exact & One-qubit & Two-qubit\\
contrast & dimension & error & infimum error\\
\midrule
$p=0$ & $2$ & $0$ & $0$\\
$0<p<1$ & $4$ & $p/2$ & $0$\\
$p=1$ & Impossible & $1/2$ & $0$ (unattained)\\
\bottomrule
\end{tabular}
\end{table}

This family also illustrates the difference between a graph and a process without using an invertible channel. For $0<p<1$, direct calculation gives
\begin{equation}
 S_{\mathcal N_{p,U}}=
 \M_2\otimes\spn\{I_b,U,U^*\}.
 \label{eq:examplegraph}
\end{equation}
Its split graph has cost $b$, whereas the process costs $2b$ for nonscalar $U$. For nonscalar $U$ at $p=1$, the off diagonal corners contain only $U$ and $U^*$, so $A$ is no longer contained in the graph. Graph monotonicity gives the same obstruction to exact synthesis as no-programming.

If the shortest spectral arc of $U$ has length $\Theta\le\pi$, then $\nu(U)=\cos(\Theta/2)$ and $e(p,U)=(p/2)\sin(\Theta/2)$, and if the spectral convex hull contains zero, $e(p,U)=p/2$. For $U_\theta=\operatorname{diag}(1,e^{i\theta})$, $-\pi\le\theta\le\pi$,
\begin{equation}
 e(p,U_\theta)=\frac p2\left|\sin\frac\theta2\right|.
 \label{eq:phase}
\end{equation}
Common output depolarization or flagged erasure of strength $\eta$ scales the branch difference and error at the minimum dimension by $1-\eta$. For nonscalar $U$ and $p=1$, depolarization with $0<\eta<1$ also restores exact cost $2b$, where the branches become full Choi rank and distinct, so Theorem~\ref{prop:controlledfeasible} and Eq.~\eqref{eq:binaryexact} apply.

For $U=Z$, identical $|+\rangle_F$ probes followed by an $X$ measurement give an expectation difference $2p$, or $2(1-\eta)p$ after depolarization. Every admissible device of minimum dimension gives zero difference. At $p=1/2$, this recovers Sec.~\ref{sec:earlyexample} and $\epsA_2=1/4$, providing a witness of retained control information.

\subsection{Amplitude damping and exact classical infeasibility}
\label{sec:damping}
The feasibility criterion also applies to dissipative noise. Let $\mathcal A_\gamma$ be the qubit amplitude damping channel with Kraus operators
\begin{equation}
 K_{\gamma,0}=\begin{pmatrix}1&0\\0&\sqrt{1-\gamma}\end{pmatrix},
 \qquad K_{\gamma,1}=\sqrt\gamma\,|0\rangle\langle1|.
 \label{eq:dampingkraus}
\end{equation}
No finite processor can program the entire amplitude damping family \cite{HilleryMaps}, and the programmability of Lindblad semigroups is studied in Ref.~\cite{JingOpenSystems}. We consider two prescribed damping parameters with the first stage constraint $S_\Phi=A$ and determine their quantum and classical carrier costs. Distinct extreme nonunitary channels can admit nonorthogonal programs \cite{Heinosaari}, so the overlap requirement does not by itself exclude exact synthesis.

\begin{corollary}[Two damping parameters]
\label{cor:damping}
Let $a=b=2$, $0<\gamma_0,\gamma_1<1$, and $\gamma_0\ne\gamma_1$. For
\begin{equation}
 \mathcal A_\oplus(X)=
 \mathcal A_{\gamma_0}(X_{00})+\mathcal A_{\gamma_1}(X_{11}),
 \label{eq:dampingtarget}
\end{equation}
the exact carrier costs and classical approximation error satisfy
\begin{equation}
 m_A(\mathcal A_\oplus)=4,\qquad
 m_A^{\rm cl}(\mathcal A_\oplus)=+\infty,\qquad
 \varepsilon_4^{A,\rm cl}(\mathcal A_\oplus)=0.
 \label{eq:dampingcosts}
\end{equation}
The last value is an unattained infimum.
\end{corollary}

\begin{proof}
Put $r=\sqrt{(1-\gamma_0)(1-\gamma_1)}$ and $d=\sqrt{\gamma_0\gamma_1}>0$. Two cross-Kraus products are
\begin{equation}
 K_{\gamma_0,0}^*K_{\gamma_1,0}=\operatorname{diag}(1,r),
 \quad K_{\gamma_0,1}^*K_{\gamma_1,1}=d|1\rangle\langle1|.
 \label{eq:dampingcross}
\end{equation}
Adding $(1-r)/d$ times the second to the first gives $I_2$. Theorem~\ref{prop:controlledfeasible} supplies a carrier of dimension four, and the distinct branches exclude dimension two.

Each $\mathcal A_\gamma$ in this range is an extreme channel since its four products $K_{\gamma,i}^*K_{\gamma,j}$ are linearly independent and thus the Kraus extremality criterion applies \cite{Choi}. For commuting programs, diagonalization gives probabilities $q_{s\ell}$ and fixed channels $\mathcal D_\ell$ with
\begin{equation}
 \mathcal A_{\gamma_s}=\sum_\ell q_{s\ell}\mathcal D_\ell.
 \label{eq:dampingmixture}
\end{equation}
Extremality forces $\mathcal D_\ell=\mathcal A_{\gamma_s}$ whenever $q_{s\ell}>0$. Since the branches differ, their classical program supports must be disjoint, contradicting Eq.~\eqref{eq:overlap}. Nevertheless, the two classical programs $(1-\epsilon,\epsilon)$ and $(\epsilon,1-\epsilon)$, with $0<\epsilon<1/2$ and decoders $\mathcal A_{\gamma_0},\mathcal A_{\gamma_1}$, have full graph and branch errors at most $\epsilon$ in normalized diamond distance. Taking $\epsilon\to0$ proves the last assertion.
\end{proof}

If one damping parameter is zero and the other is positive, the cross-Kraus span no longer contains $I_2$, so even exact quantum synthesis is impossible. The next example gives a positive gap in approximation error at a fixed carrier size.

\begin{figure*}[t]
\centering
\begin{tikzpicture}[>=Latex,font=\small]
 \begin{scope}[shift={(0,0)}]
  \node[anchor=west,font=\bfseries] at (-1.75,2.25)
       {(a) Three programs in one qubit};
  \draw[gray] (0,0.55) circle (1.12);
  \fill (0,0.55) circle (1.2pt);
  \foreach \ang/\lab in {0/1,120/2,240/3}{
    \draw[->,very thick,quantumblue] (0,0.55)
       -- ++(\ang:1.12);
    \node at ({1.44*cos(\ang)},{0.55+1.44*sin(\ang)})
       {$\rho_{\lab}$};
  }
  \node[align=center] at (0,-1.28)
       {Equatorial Bloch vectors\\
        $\Tr(\rho_j\rho_\ell)=1/4$ for $j\ne\ell$};
  \draw[->] (1.6,0.55)--(2.15,0.55);
  \node[draw,align=center,minimum width=1.65cm,
        minimum height=0.85cm] at (3.2,0.55)
       {Trine POVM\\$Q_j=2\rho_j/3$};
  \draw[->] (3.2,0.05)--(3.2,-0.48);
  \node[align=center] at (3.2,-0.97)
       {Apply $I$, $\sigma_x$, or $\sigma_z$\\on $F$};
 \end{scope}
 \begin{scope}[shift={(7.5,-0.55)}]
  \node[anchor=west,font=\bfseries] at (-0.45,2.8)
       {(b) Optimal error at fixed carrier size};
  \draw[->] (-0.1,0)--(5.1,0);
  \draw[->] (0,-0.1)--(0,2.48);
  \foreach \x/\lab in {0/2,2.2/4,4.4/6}{
    \draw (\x,0.05)--(\x,-0.05);
    \node[below] at (\x,-0.06) {$\lab$};
  }
  \draw (-0.06,2)--(0.06,2);
  \node[left] at (-0.1,2) {$1/3$};
  \draw (-0.06,1.5)--(0.06,1.5);
  \node[left] at (-0.1,1.5) {$1/4$};
  \node[rotate=90] at (-0.96,1.02) {Normalized error};
  \node at (2.35,-0.65) {Total carrier dimension $k$};
  \draw[very thick,quantumblue] (0,2)--(2.2,2);
  \draw[very thick,quantumblue] (2.2,0)--(4.88,0);
  \draw[very thick,classicalorange,dashed] (0,2)--(2.2,2);
  \draw[very thick,classicalorange,dashed] (2.2,1.5)--(4.4,1.5);
  \draw[very thick,classicalorange,dashed] (4.4,0)--(4.88,0);
  \draw[draw=quantumblue,fill=white,thick] (2.2,2) circle (2.6pt);
  \draw[draw=classicalorange,fill=white,thick] (4.4,1.5) circle (2.6pt);
  \fill[quantumblue] (0,2) circle (2.6pt);
  \fill[quantumblue] (2.2,0) circle (2.6pt);
  \fill[classicalorange] (2.2,1.5) circle (2.6pt);
  \fill[classicalorange] (4.4,0) circle (2.6pt);
  \draw[very thick,quantumblue] (0.3,0.95)--(0.85,0.95);
  \node[right] at (0.9,0.95) {Quantum register};
  \draw[very thick,classicalorange,dashed]
       (0.3,0.62)--(0.85,0.62);
  \node[right] at (0.9,0.62) {Classical register};
 \end{scope}
\end{tikzpicture}
\caption{Quantum and classical retained registers for the three-branch Pauli channel, with $\eta=0$. (a) The known trine encoding and measurement \cite{HeinosaariCommunication} provide branch probabilities $2/3,1/6,1/6$. The program of dimension two accompanies the protected qubit across the cut. (b) Exact process synthesis requires total dimension $4$ for quantum programs and $6$ for classical programs. The graph alone requires only $2$. Open endpoints indicate the value before the next dimension step, and filled points give attained errors. No device is admissible for $k<2$. Common output depolarization scales the plotted errors by $1-\eta$. The comparison excludes free shared randomness. } \label{fig:trine}
\end{figure*}
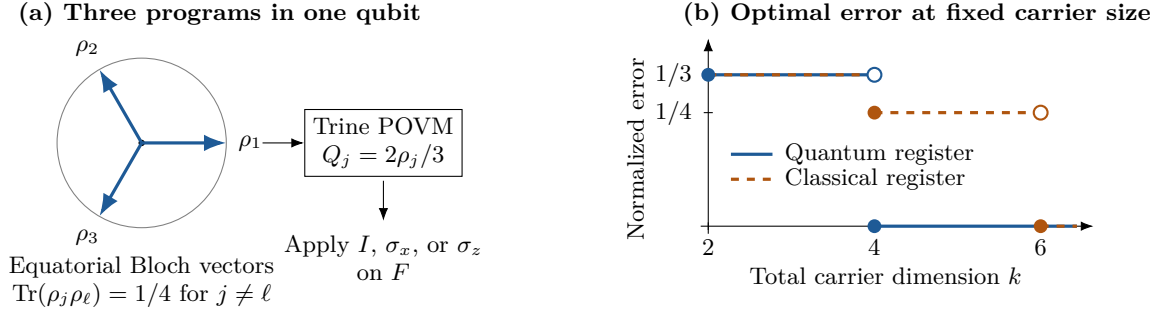

\subsection{Three branches: quantum and classical retained registers}
\label{sec:trine}
We now quantify the error imposed by retaining a classical bit instead of a qubit. The underlying trine communication matrix and its dimension separation between quantum and classical messages, $2$ versus $3$, are established \cite{HeinosaariCommunication}. We realize these probabilities as controlled Pauli channels and determine their optimal errors at fixed carrier size.

Take $a=3$, $b=2$, and put
\begin{align}
 (U_1,U_2,U_3)&=(I_2,\sigma_x,\sigma_z),\qquad
 \mathcal U_j=\Ad_{U_j},\nonumber\\
 \mathcal B_j&=\frac23\mathcal U_j+
                    \frac16\sum_{\ell\ne j}\mathcal U_\ell,\nonumber\\
 \mathcal B_\oplus(W)&=\sum_{j=1}^3\mathcal B_j(W_{jj}).
 \label{eq:trinebranches}
\end{align}
With $\omega=e^{2\pi i/3}$, use the qubit programs
\begin{equation}
 |p_j\rangle=\frac{|0\rangle+\omega^{j-1}|1\rangle}{\sqrt2},
 \quad \rho_j=|p_j\rangle\langle p_j|,
 \quad Q_j=\frac23\rho_j.
 \label{eq:trineprograms}
\end{equation}
The encoder is $\Lambda(X_E)=\sum_j\langle j|X_E|j\rangle\rho_j$. Since $\sum_j\rho_j=3I_2/2$ and $\Tr(\rho_j\rho_\ell)=1/4$ for $j\ne\ell$, it has full qubit support and graph $\M_3$. The processor
\begin{equation}
 \mathcal P(\tau\otimes X)
   =\sum_j\Tr(Q_j\tau)U_jXU_j^*
 \label{eq:trineprocessor}
\end{equation}
implements the branches: $\{Q_j\}$ is a POVM and $\Tr(Q_\ell\rho_j)$ equals $2/3$ for $\ell=j$ and $1/6$ otherwise.

\begin{proposition}[Optimal errors for trine programs]
\label{prop:trine}
Let $\mathcal R(X)=\Tr(X)I_2/2$ and replace each branch by $\mathcal B_j^{(\eta)}=(1-\eta)\mathcal B_j+\eta\mathcal R$, where $0\le\eta\le1$. The controlled target has graph $\M_6$, and $\gamma_{\mathrm{int}}(A,\M_6)=2$. For $\eta<1$,
\begin{equation}
 m_A(\mathcal B_\oplus^{(\eta)})=4,
 \qquad m_A^{\rm cl}(\mathcal B_\oplus^{(\eta)})=6.
 \label{eq:trinecosts}
\end{equation}
The optimal normalized errors are
\begin{align}
 \epsA_2(\mathcal B_\oplus^{(\eta)})
 &=\varepsilon_2^{A,\mathrm{cl}}(\mathcal B_\oplus^{(\eta)})
   =\frac{1-\eta}{3},\nonumber\\
 \epsA_4(\mathcal B_\oplus^{(\eta)})&=0,\qquad
 \varepsilon_4^{A,\mathrm{cl}}(\mathcal B_\oplus^{(\eta)})
   =\frac{1-\eta}{4}.
 \label{eq:trineerrors}
\end{align}
At dimension $6$, both errors vanish. All these optima are attained. At $\eta=1$, both exact process costs reduce to $2$.
\end{proposition}

With a classical bit, each branch must be approximated by a mixture of two fixed channels. Appendix~\ref{app:trine} proves the optimal error by reducing the distance between Pauli channels to total variation and the classical approximation problem to fitting three probability vectors by a line segment. At total carrier dimension $2$, the optimal error exceeds the diameter lower bound in Eq.~\eqref{eq:radiusbounds} for $\eta<1$.

The error advantage persists under perturbations of the target branches, while the graph constraint on the first stage remains exact.

\begin{corollary}[Stability under target perturbations]
\label{cor:trinerobust}
Let $0\le\eta<1$ and let $\widetilde{\mathcal B}_\oplus$ be a controlled process with three qubit branches satisfying
\begin{equation}
 \max_j\frac12\|\widetilde{\mathcal B}_j-
                  \mathcal B_j^{(\eta)}\|_\diamond\le\delta.
 \label{eq:trineperturb}
\end{equation}
Then
\begin{align}
 \epsA_4(\widetilde{\mathcal B}_\oplus)&\le\delta,\nonumber\\
 \varepsilon_4^{A,\mathrm{cl}}(\widetilde{\mathcal B}_\oplus)
 &\ge\max\{0,(1-\eta)/4-\delta\}.
 \label{eq:trinerobust}
\end{align}
The quantum error is strictly smaller whenever $\delta<(1-\eta)/8$.
\end{corollary}

\begin{proof}
The identity for the block norm in Appendix~\ref{app:programs} gives $\frac12\|\widetilde{\mathcal B}_\oplus-\mathcal B_\oplus^{(\eta)}\|_\diamond\le\delta$. Reusing the exact quantum realization proves the upper bound. For every admissible classical realization, the triangle inequality bounds its error on the perturbed target below by its error on the original target minus $\delta$. Taking the infimum and applying Proposition~\ref{prop:trine} gives the lower bound. Comparing the two bounds proves strict advantage in the stated range.
\end{proof}

The encoder $\Lambda$ is entanglement breaking, but its noncommuting outputs serve as quantum programs. The protected subsystem $F$ remains quantum in both comparisons. Encoding a label into a quantum state for a later measurement also underlies prepare-and-measure \cite{HeinosaariTasks,BaeExclusion} and temporal memory \cite{VieiraMemory} tasks. A free shared random seed allows convex mixtures of communication strategies \cite{deVicenteShared,FrenkelWeiner}. With this resource, Doolittle and Chitambar quantify the classical communication needed to reproduce correlations between a channel's inputs and outputs \cite{DoolittleChitambar}.

\section{Discussion}
The split criterion identifies when a target graph requires only the protected $b$-dimensional subsystem and gives a sharp $2b$ lower bound otherwise. Specifying the process can increase the memory cost by an unbounded factor, even for split graphs. For controlled channels, Theorem~\ref{prop:controlledfeasible} decides exact feasibility from the target graph, while the approximation formulas quantify the loss of control information.

Quantum programs offer two different benefits in this model. For the amplitude damping pair, they permit exact synthesis where no finite classical register suffices, but this distinction disappears at the level of infimum error. The trine example gives a positive error gap at fixed carrier size that persists under branch perturbations. Its full graph supports no nontrivial zero-error communication \cite{DSW}, so capacity alone does not determine synthesis cost. The noncommuting programs act as one way messages selecting later operations \cite{HeinosaariCommunication,HeinosaariTasks,HeYang}.

The split criterion characterizes the target graphs attainable at carrier dimension $b$, higher values of $\gamma_{\mathrm{int}}(A,T)$ for general target graphs remain to be determined. Computing the approximation error at arbitrary carrier dimensions also requires optimizing the encoder and decoder jointly. A further question is how these costs change when $F$ is only approximately correctable, so that the exact subsystem factorization must be replaced by recovery estimates \cite{BenyOreshkov}.

\section*{Data availability}
This work is analytic. No experimental or simulation data were created or analyzed. The constructions and proofs needed to reproduce the results are included in the article and its appendices.

\appendix

\section{Supported form of the protected first stage}
\label{app:factor}
We prove the subsystem factorization in Eq.~\eqref{eq:normal} \cite{NielsenPoulin,KribsSpekkens,BenyKempfKribs}, including the identification of the output support. First, support compression preserves every Kraus product. Indeed, if $y\in\ker\Phi(I)$, then $\sum_i\|V_i^*y\|^2=0$, so $V_i^*y=0$ for every $i$. Thus all Kraus ranges lie in $K=\supp\Phi(I)$, and restricting the output to $K$ changes neither the channel graph nor any subsequent composition.

Let $V_i:E\otimes F\to K$ now be Kraus operators of a fully supported channel with $S_\Phi\subseteq A$. Define the row operator
\begin{equation}
 \mathbb V:(E\otimes F)^{\oplus r}\to K,
 \qquad \mathbb V(\xi_1,\ldots,\xi_r)=\sum_iV_i\xi_i.
\end{equation}
Full support of $\Phi(I)=\sum_iV_iV_i^*$ means that $\mathbb V$ is onto. The Gram operator
\begin{equation}
 Q=\mathbb V^*\mathbb V=[V_i^*V_j]_{i,j}
\end{equation}
commutes with $\rho(x)=I_r\otimes I_E\otimes x$ for every $x\in\M_b$.

Define $\pi(x)$ on $K$ by
\begin{equation}
 \pi(x)\mathbb V\xi=\mathbb V\rho(x)\xi.
 \label{eq:rep}
\end{equation}
This is well defined, where $\ker\mathbb V=\ker Q$ is indicated by $\|\mathbb V\xi\|^2=\langle\xi,Q\xi\rangle$. Commutation therefore implies that $\rho(x)$ preserves this kernel. Hence two preimages of the same vector in $K$ give the same value in Eq.~\eqref{eq:rep}. Also,
\begin{equation}
 \|\mathbb V\rho(x)\xi\|^2
 =\|Q^{1/2}\rho(x)\xi\|^2
 \le\|x\|^2\|\mathbb V\xi\|^2.
\end{equation}
Linearity is immediate, and surjectivity gives $\pi(I)=I_K$. Multiplicativity follows by applying both sides to a spanning vector,
\begin{equation*}
 \pi(x)\pi(y)\mathbb V\xi
 =\mathbb V\rho(x)\rho(y)\xi
 =\pi(xy)\mathbb V\xi.
\end{equation*}
Commutation with $Q$ gives
\begin{align}
 \langle\mathbb V\xi,\pi(x)\mathbb V\eta\rangle
 &=\langle\xi,Q\rho(x)\eta\rangle\nonumber\\
 &=\langle\mathbb V\rho(x^*)\xi,\mathbb V\eta\rangle,
\end{align}
so $\pi(x)^*=\pi(x^*)$. To identify this representation explicitly, let $e_{\mu\nu}$ be matrix units on $F$ and set $G=\pi(e_{11})K$. Define
\begin{equation*}
 W:G\otimes F\to K,\qquad
 W(g\otimes|\mu\rangle)=\pi(e_{\mu1})g.
\end{equation*}
The identities for matrix units give $\langle\pi(e_{\mu1})g,\pi(e_{\nu1})h\rangle =\delta_{\mu\nu}\langle g,h\rangle$. Moreover, $\sum_\mu\pi(e_{\mu\mu})=I_K$, so these mutually orthogonal ranges exhaust $K$. Hence $W$ is unitary. With $U=W^*$, one has $U\pi(x)U^*=I_G\otimes x$.

Equation~\eqref{eq:rep} now implies
\begin{equation}
 (I_G\otimes x)UV_i=UV_i(I_E\otimes x).
\end{equation}
To see the tensor form, take matrix coefficients of $UV_i$ between basis vectors of $E$ and $G$. Each resulting operator on $F$ commutes with every $x\in\M_b$, and therefore is a scalar multiple of $I_F$. Collecting these scalars gives $UV_i=C_i\otimes I_F$. Trace preservation yields $\sum_iC_i^*C_i=I_E$, so the $C_i$ define a channel $\Lambda$. Their products give $S_\Phi=S_\Lambda\otimes\C I_b$. Finally, $U\Phi(I)U^*=\Lambda(I_E)\otimes I_F$ is strictly positive on $G\otimes F$. Thus $\Lambda$ has full support and the supported dimension is exactly $b\dim G$.

\section{Graph realization and the split criterion}
\label{app:graph}

\subsection{Universal feasibility}
Choose a channel $\Phi_0$ of width $\gamma(S)$ with graph $S$, and a channel $\Psi$ with graph $T$. The flagged map
\begin{equation}
 \widehat\Phi(X)=\tfrac12\Phi_0(X)\oplus\tfrac12X
\end{equation}
has graph $S+\C I_n=S$ and width $\gamma(S)+n$. Its Kraus operators are those of $\Phi_0$, embedded in the first output sector, together with the identity embedded in the second, all scaled by $1/\sqrt2$. Products between the two sectors vanish, which explains the sum of graphs. Let the postprocessor discard off diagonal flag blocks and act as the identity on the first sector and as $\Psi$ on the second, retaining orthogonal output flags. Then
\begin{equation}
 (\widehat{\mathcal D}\circ\widehat\Phi)(X)
   =\tfrac12\Phi_0(X)\oplus\tfrac12\Psi(X).
\end{equation}
Its graph is $S+T=T$. This proves Eq.~\eqref{eq:universal}. The construction counts both flag sectors as part of the carrier.

\subsection{The minimum carrier and matrix factors}
For the trace channel on $E$, use Kraus operators $W_s=\langle s|\otimes I_b$. If $B_\alpha$ are Kraus operators of $\mathcal D$, then
\begin{equation}
 (B_\alpha W_s)^*(B_\beta W_t)
 =|s\rangle\langle t|\otimes B_\alpha^*B_\beta.
\end{equation}
This proves Eq.~\eqref{eq:tracegraph}. Equation~\eqref{eq:normalgraph} excludes any supported carrier below $b$ and shows that $b$ is attainable for $A$. Hence the graph criterion at the minimum dimension follows in both directions.

To verify the equivalent multiplication criterion, write $t=\sum_{p,q}|p\rangle\langle q|\otimes t_{pq}\in T$. If $ATA\subseteq T$, then
\begin{equation}
 (|i\rangle\langle p|\otimes I)t
 (|q\rangle\langle j|\otimes I)
 =|i\rangle\langle j|\otimes t_{pq}\in T.
\end{equation}
Let $R$ be the span of all coefficients $t_{pq}$ of all elements of $T$. This identity gives $\M_a\otimes R\subseteq T$, and the reverse containment follows from the coefficient expansion. Since $T$ contains $I_{ab}$ and is self adjoint, $R$ contains $I_b$ and is self adjoint. The converse is immediate. The space $R$ is unique, being recoverable from the coefficient slices. Integrality of the supported dimensions and the universal construction complete Theorem~\ref{thm:split}.

\subsection{The sharp nonsplit realization}
For the Kraus family in Eq.~\eqref{eq:sharpC},
\begin{equation*}
 \sum_iC_i^*C_i=I_a,\qquad
 \Lambda(I_a)=(a-p)|0\rangle\langle0|+p|1\rangle\langle1|.
\end{equation*}
Both output coefficients are positive. Put $d_1=\sqrt{1-p}$ and $d_r=1$ for $r\ge2$. For $1\le r,s\le a$, the operators with output $|0\rangle$ give $C_r^*C_s=d_rd_s|r\rangle\langle s|$. Since all these coefficients are nonzero, their products span $\M_a$.

Let $P=|1\rangle\langle1|_G$. The second generator of $R_Z$ has pullback terms
\begin{equation*}
 (C_i^*\otimes I_b)(P\otimes Z)(C_j\otimes I_b)
 =C_i^*PC_j\otimes Z.
\end{equation*}
All vanish except $i=j=0$, which gives $p|1\rangle\langle1|_E\otimes Z$. The identity generator supplies exactly $A$, so no additional graph directions occur. Because $Z$ is nonscalar, this new direction is linearly independent of $A$. Thus $\dim T_Z=a^2+1$, proving nonsplitting and the sharp cost $2b$.

The postprocessor can also be made explicit. Put $H=P\otimes Z$, choose $0<t<\|H\|^{-1}$, and define $M_\pm=(I_{2b}\pm tH)/2$. The channel
\begin{equation}
 \mathcal D(Y)=\sqrt{M_+}Y\sqrt{M_+}
                   \oplus\sqrt{M_-}Y\sqrt{M_-}
 \label{eq:sharpdecoder}
\end{equation}
is trace preserving because $M_++M_-=I_{2b}$. Its two Kraus operators have orthogonal output ranges, so their cross products vanish and $S_{\mathcal D}=\spn\{M_+,M_-\}=R_Z$. The output flag is created after the intermediate cut and therefore does not contribute to the carrier dimension.

\section{Tests for the minimum carrier and linear bottlenecks}
\label{app:bottleneck}

\subsection{Erasure invariance and its error bounds}
\label{app:erase}
At dimension $b$, Eq.~\eqref{eq:minimumtrace} gives $\mathcal N=\mathcal D\circ\Tr_E$. Such a channel satisfies $\mathcal N=\mathcal N\circ\mathcal E_E$. Conversely, if this invariance holds, the channel $\mathcal D(Y)=\mathcal N(I_E/a\otimes Y)$ supplies the required factorization. This proves Eq.~\eqref{eq:erasetest}.

For the upper bound in Eq.~\eqref{eq:erasebound}, use $\mathcal N\circ\mathcal E_E$ as the approximant. For the lower bound, any $\mathcal M\in\CA(b)$ obeys $\mathcal M=\mathcal M\circ\mathcal E_E$, whence
\begin{align*}
 \mathcal N-\mathcal N\circ\mathcal E_E
 &=(\mathcal N-\mathcal M)
       -(\mathcal N-\mathcal M)\circ\mathcal E_E,\\
 d_E(\mathcal N)&\le2\|\mathcal N-\mathcal M\|_\diamond.
\end{align*}
The last inequality uses $\|\mathcal E_E\|_\diamond=1$. Taking the infimum over $\mathcal M$ and inserting the normalization in Eq.~\eqref{eq:error} gives the factor $1/4$.

\subsection{Lower bounds from rank and the norm of the inverse}
If $\mathcal M$ factors through $\M_k$, then its image has linear dimension at most $\dim\M_k=k^2$. For the invertible target in Proposition~\ref{prop:invertible}, this proves $m_A(\mathcal N)\ge ab$. Invertibility also implies full support of $\mathcal N(I)$. To see this, for every $X\ge0$ one has $0\le\mathcal N(X)\le\|X\|\mathcal N(I)$, so $\mathcal N(X)$ is supported on $\supp\mathcal N(I)$. Positive matrices linearly span the input algebra. A proper output support would therefore constrain the entire range of $\mathcal N$ to a proper matrix subspace, contradicting surjectivity. Taking $\Phi=\mathcal N$ and $\mathcal D=\id$ proves equality.

When $k<ab$, a factorized $\mathcal M$ has a nonzero Hermitian kernel element. Indeed, its kernel is closed under adjoints, so the Hermitian or anti-Hermitian part of any nonzero kernel element supplies one. Normalize it to $\|H\|_1=1$.
Then
\begin{align}
 1&=\|\mathcal N^{-1}(\mathcal N(H))\|_1
 \le\|\mathcal N^{-1}\|_\diamond\|\mathcal N(H)\|_1,\nonumber\\
 \|\mathcal N-\mathcal M\|_\diamond
 &\ge\|\mathcal N(H)\|_1.
\end{align}
The diamond norm bounds the induced trace norm even on a nonpositive input, so these inequalities prove Eq.~\eqref{eq:inverse}.

For Eq.~\eqref{eq:depolbound}, use the protected form of the actual first stage. If $k<ab$, the supported dimension of its retained register satisfies $c<a$. Since $\dim\M_c<\dim\M_a$, choose Hermitian $H\ne0$ with $\Lambda(H)=0$. Trace preservation gives $\Tr H=0$. For $\|H\|_1=1$ and a state $\tau$ on $F$, the admissible composite annihilates $H\otimes\tau$, whereas $\mathcal N_q(H\otimes\tau)=(1-q)H\otimes\tau$. Here $\|H\otimes\tau\|_1=1$ because the trace norm is multiplicative on tensor products. The diamond distance is therefore at least $1-q$ for each admissible composite. The choice of $H$ may depend on that composite, but the bound does not. Taking the infimum and the normalization factor $1/2$ proves Eq.~\eqref{eq:depolbound}.

\section{Closure of the protected realization set}
\label{app:closure}
Fix $c$ and the final output space. Choi matrix of each channel is positive with a prescribed partial trace, hence has a fixed trace. The channel sets are therefore closed and bounded in finite-dimensional matrix spaces, and thus compact. Composition is continuous, so their image $\mathcal F_c$ is compact as well.

An element of $\CA(cb)$ has a supported form whose retained register has dimension $c'\le c$. Let $J:\C^{c'}\to\C^c$ be an isometry, put $W=J\otimes I_b$ and $P=WW^*$, and replace the encoder by $\widetilde\Lambda(X)=J\Lambda(X)J^*$. If $\mathcal D'$ is the original decoder, a channel extending it is
\begin{equation*}
 \widetilde{\mathcal D}(Y)
 =\mathcal D'(W^*YW)+\Tr[(I-P)Y]\sigma,
\end{equation*}
where $\sigma$ is any fixed output state. Both terms are completely positive, and their traces sum to $\Tr Y$. The second term vanishes on the encoded output, while the first reproduces the original decoder. Thus the composite is unchanged. This proves $\CA(cb)\subseteq\mathcal F_c$ and, since $\mathcal F_c$ is closed, $\overline{\CA(cb)}\subseteq\mathcal F_c$.

Conversely, take any $\mathcal D\circ(\Lambda\otimes\id_b)\in\mathcal F_c$ and use $\Lambda_\delta$ from Eq.~\eqref{eq:fulldense}. A Kraus family for the completely depolarizing component is $c^{-1/2}|r\rangle\langle s|$, $1\le r\le c$, $1\le s\le a$. Products with the same $r$ span all matrix units $|s\rangle\langle t|$, so $S_{\Lambda_\delta}=\M_a$. Also $\Lambda_\delta(I_a)\ge\delta aI_c/c$, proving full support. Therefore the perturbed composite belongs to $\CA(cb)$. Contraction of the diamond norm under channels gives
\begin{align}
 &\frac12\|\mathcal D\circ(\Lambda_\delta\otimes\id_b)
       -\mathcal D\circ(\Lambda\otimes\id_b)\|_\diamond\nonumber\\
 &\hspace{3em}\le\tfrac12\|\Lambda_\delta-\Lambda\|_\diamond\le\delta.
\end{align}
This proves the reverse inclusion in Eq.~\eqref{eq:closure}. Finally, the distance to a fixed target is continuous. Every point of $\mathcal F_c$ has constrained approximants with converging distances, so passing to the closure does not change the infimum. Compactness of $\mathcal F_c$ makes that infimum a minimum, proving Eq.~\eqref{eq:relaxedmin} without asserting attainment in $\CA(cb)$ itself.

\section{Controlled channels and programming}
\label{app:programs}

\subsection{Exact synthesis}
Suppose $\mathcal N_\oplus=\mathcal P\circ(\Lambda\otimes\id_b)$ is an admissible supported factorization. Put $\rho_s=\Lambda(|s\rangle\langle s|)$. Evaluating the factorization on $|s\rangle\langle s|\otimes X$ gives Eq.~\eqref{eq:programs}. For the overlap identity, write $u_{is}=C_i|s\rangle$, so $\rho_s=\sum_i|u_{is}\rangle\langle u_{is}|$. The trace of the product of two such rank-one operators is $|\langle u_{is},u_{jt}\rangle|^2$. Summing proves Eq.~\eqref{eq:overlapidentity}. Every summand is nonnegative. If the sum vanished for a pair $s,t$, every Kraus product would have zero $(s,t)$ entry, contradicting $S_\Lambda=\M_a$. Thus the overlap condition is necessary.

Conversely, suppose the program states and processor satisfy Eqs.~\eqref{eq:programs} and \eqref{eq:overlap}. For the encoder
\begin{equation}
 \Lambda_\rho(X)=\sum_s\langle s|X|s\rangle\rho_s,
\end{equation}
write $\rho_s=\sum_\alpha\lambda_{s\alpha}|v_{s\alpha}\rangle\langle v_{s\alpha}|$, keeping only positive eigenvalues. Kraus operators are $C_{s\alpha}=\sqrt{\lambda_{s\alpha}}|v_{s\alpha}\rangle\langle s|$, with products
\begin{equation}
 C_{s\alpha}^*C_{t\beta}
 =\sqrt{\lambda_{s\alpha}\lambda_{t\beta}}
 \langle v_{s\alpha}|v_{t\beta}\rangle|s\rangle\langle t|.
\end{equation}
A nonzero product exists for every pair exactly when $\Tr(\rho_s\rho_t)>0$. Hence $S_{\Lambda_\rho}=\M_a$ and the composite is the required controlled channel. The supported program space is the span of the supports of the $\rho_s$ and has dimension at most $c$. This proves Proposition~\ref{thm:programs}.

The no-programming restriction also applies to mixed programs. If a mixed program gives a unitary channel, each pure state in its spectral decomposition gives the same channel, since unitary channels are extreme points of the channel set. Pure programs for distinct unitaries are orthogonal by Ref.~\cite{NielsenChuang}. Applying this to every pair of eigenvectors gives orthogonal supports for the mixed programs, as discussed in Ref.~\cite{Heinosaari}.

\subsection{A graph criterion for exact controlled synthesis}
\label{app:feasible}
The Kraus operators of the controlled channel are $\langle s|\otimes K_{s\mu}$. Their products give the independent matrix corners
\begin{equation}
 S_{\mathcal N_\oplus}
 =\sum_{s,t}E_{st}\otimes
       \spn\{K_{s\mu}^*K_{t\nu}:\mu,\nu\}.
\end{equation}
Thus graph containment is equivalent to the final condition in Eq.~\eqref{eq:controlledfeasible}. Its necessity for exact realization follows from Eq.~\eqref{eq:pullback}.

Suppose the condition holds, and let $r_s$ be the number of Kraus operators of branch $s$. For every $s<t$, choose a matrix $Q_{st}:\C^{r_t}\to\C^{r_s}$ satisfying
\begin{equation}
 \sum_{\mu,\nu}(Q_{st})_{\mu\nu}
            K_{s\mu}^*K_{t\nu}=I_b.
 \label{eq:crossidentity}
\end{equation}
Put $Q_{ts}=Q_{st}^*$. Form the Hermitian block matrix $H$ on $\bigoplus_s\C^{r_s}$ with zero diagonal blocks and off diagonal blocks $Q_{st}$. Choose $0<\alpha<1/(1+\|H\|)$. Then $B=I+\alpha H$ is strictly positive. Factoring $B=J^*J$, and denoting the column blocks by $J_s$, gives isometries into a common environment $\mathcal E$ with
\begin{equation}
 J_s^*J_s=I_{r_s},\qquad
 J_s^*J_t=\alpha Q_{st}\quad(s\ne t).
\end{equation}
Taking the adjoint of Eq.~\eqref{eq:crossidentity} verifies the same identity with $s,t$ reversed, so these choices are consistent. Positivity follows from $B\ge(1-\alpha\|H\|)I>0$ with no positivity of an individual $Q_{st}$ is required. One may take $\mathcal E=\bigoplus_s\C^{r_s}$ and $J=B^{1/2}$. Define Stinespring isometries
\begin{equation}
 W_sx=\sum_\mu K_{s\mu}x\otimes|\mu\rangle,
 \qquad V_s=(I_L\otimes J_s)W_s.
\end{equation}
Equations~\eqref{eq:crossidentity} and trace preservation imply
\begin{equation}
 V_s^*V_t=
 \begin{cases}
 I_b,&s=t,\\
 \alpha I_b,&s\ne t.
 \end{cases}
 \label{eq:Vgram}
\end{equation}
The isometric embedding $J_s$ preserves the environmental inner products, so tracing out $\mathcal E$ recovers the original branch $\mathcal N_s$.

The scalar matrix $(1-\alpha)I_a+\alpha\boldsymbol1\boldsymbol1^*$ has eigenvalues $1-\alpha$ and $1+(a-1)\alpha$, so it is positive definite. A square root supplies linearly independent unit vectors $|p_s\rangle$ in $\C^a$ with this Gram matrix. Define $V:\C^a\otimes F\to L\otimes\mathcal E$ by
\begin{equation}
 V(|p_s\rangle\otimes x)=V_sx
\end{equation}
and extend linearly. Linear independence makes this definition well defined on the full domain. More explicitly, for arbitrary $x_s\in F$, Eq.~\eqref{eq:Vgram} gives
\begin{align*}
 \left\|\sum_sV_sx_s\right\|^2
 &=\sum_{s,t}\langle x_s,V_s^*V_tx_t\rangle\\
 &=\left\|\sum_s|p_s\rangle\otimes x_s\right\|^2.
\end{align*}
Thus $V$ is an isometry. The channel $\mathcal P(Y)=\Tr_{\mathcal E}(VYV^*)$ implements branch $s$ on program $|p_s\rangle\langle p_s|$. These programs span $\C^a$, so the sum of their density matrices is strictly positive. Their distinct pairwise overlaps are $\alpha^2>0$. Proposition~\ref{thm:programs} therefore supplies a fully supported carrier of dimension $ab$, proving Theorem~\ref{prop:controlledfeasible}.

\subsection{Approximate synthesis and the block norm}
Let $\Delta_s$ be Hermiticity preserving maps, and define $\Delta(X)=\sum_s\Delta_s(X_{ss})$. Then
\begin{equation}
 \|\Delta\|_\diamond=\max_s\|\Delta_s\|_\diamond.
 \label{eq:blocknorm}
\end{equation}
Inputs supported in a single block give the lower bound. To see this, fix a control label and choose a joint state of the input and reference systems that attains the corresponding branch diamond norm. For the upper bound, append an arbitrary reference $R$ and write $X_{ss}$ for the diagonal blocks of an input operator. Pinching is an average of unitary conjugations, so it contracts the trace norm even when the input is not Hermitian. Consequently,
\begin{align*}
 \|(\Delta\otimes\id_R)(X)\|_1
 &\le\sum_s\|(\Delta_s\otimes\id_R)(X_{ss})\|_1\\
 &\le\max_s\|\Delta_s\|_\diamond\sum_s\|X_{ss}\|_1\\
 &\le\max_s\|\Delta_s\|_\diamond\|X\|_1.
\end{align*}
The last step uses that the trace norm of the pinched, block diagonal operator is the sum of its block norms. Taking the supremum over inputs and reference systems proves Eq.~\eqref{eq:blocknorm}.

Let $\Delta_E$ be dephasing in the control basis. The target satisfies $\mathcal N_\oplus\circ(\Delta_E\otimes\id_b)=\mathcal N_\oplus$. For any approximate factorization in $\mathcal F_c$, precomposing the encoder with $\Delta_E$ cannot increase the error. The resulting encoder is measure-and-prepare with program states $\rho_s=\Lambda(|s\rangle\langle s|)$. Conversely, every such family is an encoder into $\M_c$. Equation~\eqref{eq:blocknorm} and Proposition~\ref{thm:closure} now prove Eq.~\eqref{eq:programapprox}. Compactness of the sets of program states and processors ensures the displayed minimum exists.

\subsection{Graph and boundary of the binary random unitary example}
Write $q_0=(1+p)/2$ and $q_1=(1-p)/2$. A Kraus family for Eq.~\eqref{eq:example} is
\begin{equation}
 K_{s,0}=\sqrt{1-q_s}\,\langle s|\otimes I_b,\qquad
 K_{s,1}=\sqrt{q_s}\,\langle s|\otimes U,
 \label{eq:exampleK}
\end{equation}
for $s=0,1$. When $0\le p<1$, all four coefficients are positive. For every matrix corner $E_{st}$, their products span $E_{st}\otimes\spn\{I_b,U,U^*\}$. This proves Eq.~\eqref{eq:examplegraph}. At $p=1$, only $\langle0|\otimes U$ and $\langle1|\otimes I_b$ survive, and the graph becomes
\begin{equation}
 \spn\{E_{00}\otimes I_b,E_{11}\otimes I_b,
                E_{01}\otimes U^*,E_{10}\otimes U\}.
 \label{eq:boundarygraph}
\end{equation}
For nonscalar $U$, it cannot contain $E_{01}\otimes I_b$, so it does not contain $A$. This gives the direct graph obstruction at $p=1$. Meanwhile,
\begin{equation}
 \|\mathcal N_{1,U}-\mathcal N_{p,U}\|_\diamond
 =\frac{1-p}{2}\|\Ad_U-\id\|_\diamond\longrightarrow0,
\end{equation}
by Eq.~\eqref{eq:blocknorm}. Each approximating channel has an admissible $2b$-dimensional realization, which proves the nonattainment assertion.

\section{Radius at the minimum dimension and its semidefinite form}
\label{app:radius}
At dimension $b$, all admissible processes are $\mathcal D\circ\Tr_E$. Equation~\eqref{eq:blocknorm} therefore proves Eq.~\eqref{eq:radius}. For any center $\mathcal D$, the triangle inequality gives
\begin{equation}
 \|\mathcal N_s-\mathcal N_t\|_\diamond
 \le\|\mathcal N_s-\mathcal D\|_\diamond+
       \|\mathcal N_t-\mathcal D\|_\diamond.
\end{equation}
This is the lower bound in Eq.~\eqref{eq:radiusbounds}. For the upper bound, use $\overline{\mathcal N}=a^{-1}\sum_t\mathcal N_t$ and
\begin{equation}
 \|\mathcal N_s-\overline{\mathcal N}\|_\diamond
 \le\frac1a\sum_{t\ne s}\|\mathcal N_s-\mathcal N_t\|_\diamond.
\end{equation}
For $a=2$ both bounds coincide and the average is a midpoint, proving Eq.~\eqref{eq:binary}.

For the semidefinite formulation, use the unnormalized Choi matrix with the output factor first, $J(\mathcal L)=\sum_{i,j}\mathcal L(|i\rangle\langle j|)\otimes|i\rangle\langle j|$. A decoder is a channel precisely when $J_D\ge0$ and $\Tr_LJ_D=I_F$.

For a Hermiticity preserving map $\Delta$, the diamond norm dual program
\cite{WatrousSDP} specializes to
\begin{equation*}
 \|\Delta\|_\diamond
 =\min_{Z\ge J(\Delta),\ Z\ge-J(\Delta)}
       \|\Tr_LZ\|_\infty.
\end{equation*}
To see why one matrix suffices, the general dual uses a positive block matrix with diagonal variables $Y_0,Y_1$ and off diagonal entries $-J(\Delta)$. Hermiticity permits exchanging the diagonal variables. Averaging the two feasible block matrices gives equal diagonals $Z=(Y_0+Y_1)/2$ without increasing the objective. Diagonalizing this $2\times2$ block structure gives precisely $Z\ge\pm J(\Delta)$ \cite{WatrousSDP}.

Apply this identity to $\Delta_s=\mathcal N_s-\mathcal D$, whose Choi matrix is $J(\mathcal N_s)-J_D$. Introduce a Hermitian matrix $Z_s$ on $L\otimes F$ for each branch and a common real bound $t$. Then twice the error at the minimum dimension is the optimal value of
\begin{align}
 &\text{minimize }t,\nonumber\\
 &J_D\ge0,\qquad\Tr_LJ_D=I_F,\nonumber\\
 &Z_s\ge J(\mathcal N_s)-J_D,\nonumber\\
 &Z_s\ge-J(\mathcal N_s)+J_D,\nonumber\\
 &\Tr_LZ_s\le tI_F\qquad\text{for all }s.
 \label{eq:sdp}
\end{align}
Adding the two matrix inequalities gives $Z_s\ge0$, hence $\Tr_LZ_s\ge0$. The last constraint is therefore equivalent to $\|\Tr_LZ_s\|_\infty\le t$. The norm dual proves both directions: every feasible tuple bounds every branch distance by $t$, while each decoder admits certificates at its largest branch distance. The minimum is consequently the radius in Eq.~\eqref{eq:radius}, or twice $\epsA_b(\mathcal N_\oplus)$.

\section{Trine programs and the classical carrier bound}
\label{app:trine}
First, a classical output encoder has the form $\Lambda(X_E)=\sum_\ell\Tr(M_\ell X_E)|\ell\rangle\langle\ell|$ for a POVM $\{M_\ell\}$. Dephasing $E$ gives commuting programs $\rho_j=\sum_\ell q_{j\ell}|\ell\rangle\langle\ell|$. If $S_\Lambda=\M_3$, its output states for any pair of basis labels have nonorthogonal supports, by Eq.~\eqref{eq:overlapidentity}. Consequently the dephased encoder still has graph $\M_3$. Precomposition with this dephasing cannot increase distance from the controlled target. It is therefore sufficient to consider these classical programs, for exact synthesis as well as approximation. For a fixed processor,
\begin{equation}
 \mathcal P(\rho_j\otimes\cdot)
       =\sum_{\ell=1}^c q_{j\ell}\mathcal D_\ell,
 \quad \mathcal D_\ell(X)=\mathcal P(|\ell\rangle\langle\ell|\otimes X).
 \label{eq:trineclassicalspan}
\end{equation}
The channels $\mathcal D_\ell$ may act arbitrarily on the protected input, so the dimension bound below applies to every such processor.

The three channels in Eq.~\eqref{eq:trinebranches} are linearly independent since their coefficients on the independent Pauli channels $\mathcal U_1,\mathcal U_2,\mathcal U_3$ form $I_3/2+\mathbf1\mathbf1^{\mathsf T}/6$, which is invertible. Their noisy versions remain independent for $\eta<1$. Indeed, a linear relation has coefficients summing to zero by trace preservation, after which its common noise and common branch average cancel. Equation~\eqref{eq:trineclassicalspan} then requires $c\ge3$. Conversely, the positive distributions $(2/3,1/6,1/6)$ and their permutations, decoded by $(1-\eta)\mathcal U_\ell+\eta\mathcal R$, give a classical realization with full support and full graph at $c=3$. Equations~\eqref{eq:trineprograms}--\eqref{eq:trineprocessor} give a quantum realization with $c=2$, and distinct branches rule out the case of $c=1$.

All coefficients of $I_2,\sigma_x,\sigma_z$ are positive in every branch, including after adding noise. The cross-Kraus products of any two branches span $\M_2$, since these Pauli products also include a nonzero multiple of $\sigma_y$. The controlled graph is therefore $\M_3\otimes\M_2=\M_6$. The split criterion gives its graph cost $2$.

It remains to prove the claims about the errors. Applying Pauli twirling to a candidate decoder cannot increase any branch error because the target branches are Pauli channels. For a classical register, twirling each $\mathcal D_\ell$ preserves Eq.~\eqref{eq:trineclassicalspan}, and this is a fixed modified processor and requires no shared random seed. For two Pauli channels, diamond distance equals the $\ell_1$ distance between their probability vectors \cite{Sacchi}: the triangle inequality gives the upper bound, and a maximally entangled probe gives the matching lower bound through orthogonal Bell outputs. Order the outcomes as $I_2,\sigma_x,\sigma_z,\sigma_y$. The target vectors are $P_1=(2/3,1/6,1/6,0)$, with $P_j$ obtained by placing $2/3$ in coordinate $j$ among the first three entries. With noise,

\begin{equation}
 P_j^{(\eta)}=(1-\eta)P_j+\frac\eta4\mathbf1.
\end{equation}

We use total variation $\operatorname{TV}(P,Q)=\tfrac12\|P-Q\|_1$, which equals the normalized diamond error for these Pauli channels. For lower bounds, apply a stochastic map that leaves the first three outcomes unchanged and distributes the fourth uniformly among them. It does not increase total variation distance, maps line segments to line segments, and sends the target distributions to
\begin{equation}
 \widetilde P_j=\frac{1-\eta}{2}e_j+
                       \frac{1+\eta}{6}\mathbf1\in\mathbb R^3.
 \label{eq:trinetriangle}
\end{equation}
For a single center $q$, some $q_j\le1/3$. Its total variation distance from $\widetilde P_j$ is at least $(2-\eta)/3-1/3=(1-\eta)/3$. The average of the three original noisy Pauli channels attains this bound.

With two classical symbols, all three approximants lie on a line segment in the probability simplex. Choose a nonconstant normal $n\in\mathbb R^3$ to its affine line, so $n\cdot q_j=h$ for every approximant. The same argument applies if the segment is a point. Put $r=\max_i n_i-\min_i n_i>0$. If every total variation error is at most $\delta$, then
\begin{equation}
 |n\cdot\widetilde P_j-h|\le r\delta.
\end{equation}

The range of $n\cdot\widetilde P_j$ over the three targets equals $(1-\eta)r/2$, proving $\delta\ge(1-\eta)/4$. For the matching construction at $\eta=0$, use $\mathcal D_1=\mathcal U_1$ and $\mathcal D_2=(\mathcal U_2+\mathcal U_3)/2$, with program distributions

\begin{equation}
 q_1=(2/3,1/3),\qquad q_2=q_3=(1/6,5/6).
\end{equation}

Branch 1 is exact, branches 2 and 3 both have distribution $(1/6,5/12,5/12,0)$ and error $1/4$. Every program has full support, so the first stage graph is exactly $A$. Replacing each decoder channel by $(1-\eta)\mathcal D_\ell+\eta\mathcal R$ scales both errors by $1-\eta$. This proves attainment and Eq.~\eqref{eq:trineerrors}.


\makeatletter
\let\auto@bib@innerbib\@empty
\makeatother

\end{document}